\documentclass[twocolumn,showpacs]{revtex4}

\usepackage{graphicx}%
\usepackage{dcolumn}
\usepackage{amsmath}

\begin{document}

\title{Green's function technique for a two-electrode mesoscopic system under bias}

\author{Jongbae Hong}

\address{Department of Physics,
Seoul National University, Seoul 151-747, Korea \\
Max-Planck-Institut f\"ur Physik Komplexer Systeme, D-01187
Dresden, Germany}

\begin{abstract}

We present a Green's function technique for studying the nonlinear
conductance of a nanocontact system with two electrodes at
different chemical potentials. The retarded Green's function for a
single-impurity Anderson model with two reservoirs is obtained in
terms of a $5\times 5$ matrix in which the effect of bias is
contained. A complete set of basis vectors for the single-impurity
Anderson model has been provided before formulating the Green's
function. Finally, we present a self-consistent method to fix the
undetermined quantities existing in the matrix elements for the
retarded Green's function.

\end{abstract}



\maketitle

\section{introduction}
The single-particle Green's function is a basic tool for studying
the correlation effects in many-body systems. However, calculation
of this function in the case of a strongly correlated system is
not usually successful because such a calculation generally
requires nonperturbative treatment. Moreover, the possibility of a
successful calculation becomes even less when the system is under
steady-state nonequilibrium (SSN) conditions. Recently, some
mesoscopic systems having nanocontact features have attracted
considerable interest. Such systems include a quantum point
contact~\cite{cronen,dicarlo,sfigakis,sarkozy}, a single electron
transistor with a quantum dot~\cite{cronen2,kogan,amasha,nygard},
and magnetic atom adsorbed on a metallic
surface~\cite{mad,mano,neel,otte}. Experimentalists measure the
$dI/dV$ in these systems for a given $V$; $I$ and $V$ denote the
current and the bias, respectively. Because of the bias, resonant
tunneling occurs unidirectionally in the ground state. This
unidirectional tunneling causes the system to be out of the linear
response regime. Therefore, nonlinear conductances observed for
the abovementioned systems may not be explained by a linear
response theory. The purpose of this study is to provide a Green's
function technique that is suitable for studying the nonlinear
conductance of a mesoscopic system operating under bias.

A well-known formula for a steady current passing through a small
interacting region that is connected to charge reservoirs having
different chemical potentials was proposed more than a decade
ago~\cite{meir,hersh}. This formula is given by
\begin{eqnarray} I&=&-\frac{e}{\hbar}
\int\frac{d\omega}{2\pi}\frac{\Gamma^L(\omega)
\Gamma^R(\omega)}{\Gamma^L(\omega)+\Gamma^R(\omega)}
[f_L(\omega)-f_R(\omega)]\nonumber
\\
&\times& {\rm Im}G_{dd\sigma}^+(\omega; V) \label{current}
\end{eqnarray}
for proportional lead functions, i.e.,
$\Gamma^L(\omega)\propto\Gamma^R(\omega)=
2\pi\sum_\sigma\rho_\sigma^R(\omega)|V^R_{kd}|^2$, where $\sigma$
indicates spin, $L$ and $R$ denote the left and right reservoirs,
respectively, $\rho_\sigma^R(\omega)$ is the density of states of
the right metallic reservoir without inter-electron interaction,
and $V^R_{kd}$ is the strength of hybridization between the state
$d$ of interacting region and the state $k$ of the right
reservoir. Further, $f_L(\omega)$ in Eq. (\ref{current}) denotes
the Fermi distribution function of the left reservoir. Since
unidirectional tunneling at ground state is a unique
characteristic of the SSN, the effect of bias must be appeared
explicitly in the retarded Green's function in Eq.
(\ref{current}). However, to the best of our knowledge the
steady-state Green's function for the quantum impurity system has
not been obtained yet.

The purpose of this study is to present a method for determining
the steady-state retarded Green's function appearing in Eq.
(\ref{current}). To realize this method, we adopt the retarded
Green's function expressed in resolvent operator form in the
Heisenberg picture~\cite{fulde}; this function is given by
\begin{equation}
iG^{+}_{ij\sigma}(z)=\langle c_{i\sigma}|(z{\bf I}+i{\rm\bf
L})^{-1}|c_{j\sigma}\rangle, \label{heisenberg}
\end{equation}
where $z=-i\omega+\eta$ and $c_{i\sigma}$ indicates annihilation
of a fermion with spin $\sigma$ at a state $i$. The symbols
${\rm\bf I}$ and ${\rm\bf L}$ in Eq. (\ref{heisenberg}) denote the
identity operator and the Liouville operator, respectively. The
latter is defined by ${\rm\bf L}{\hat{\cal O}}=[{\cal
H},{\hat{\cal O}}]={\cal H}{\hat{\cal O}}-{\hat{\cal O}}{\cal H}$,
where $\cal H$ and $\hat{\cal O}$ are the Hamiltonian operator and
an arbitrary operator, respectively. Although we are familiar with
the resolvent form in the Schr\"odinger picture,
\begin{equation}
G^{+}_{ij}(\omega)=\langle\phi_i|(\omega+i\eta-{\cal
H})^{-1}|\phi_j\rangle, \label{resolvent}
\end{equation}
where $\phi_j$ is an eigenstate of ${\cal H}$, we employ Eq.
(\ref{heisenberg}) in this study because a dynamical approach
using operators is more appropriate for describing electron
hopping under steady-state conditions.

This paper is composed as follows: In section II, we present a
systematic method to determine a complete set of basis vectors for
calculating the retarded Green's function given in Eq.
(\ref{heisenberg}). A single-impurity Anderson model with one- and
two-reservoir is studied as an example. In section III, we express
the retarded Green's function for a two-reservoir system under
bias in terms of a finite-dimensional matrix and discuss the
physical meanings of the matrix elements. We present a
self-consistent method to determine unknown quantities in section
IV and finally give conclusions in section V.

\section{Determining Complete Set of Basis Vectors}
The first step in the calculation of the retarded Green's function
in resolvent form, Eq. (\ref{heisenberg}), is ensuring a complete
set of basis vectors. It should be noted that the resolvent form
does not provide information about the orthogonal basis vectors.
Therefore, the Lanczos algorithm~\cite{lanczos,dago} in the
Sch\"odinger picture and the projection operator
technique~\cite{mori,zwan} in the Heisenberg picture are used to
obtain static and dynamic orthogonal basis vectors, respectively.
However, employing the Gram-Schmidt orthogonalization procedure to
obtain orthogonal basis vectors for a nontrivial system causes
significant complications. Hence, we propose a new and simpler
technique for obtaining the basis vectors.

The retarded Green's function in Eq. (\ref{heisenberg}) can be
obtained by calculating the matrix inverse $({\rm\bf
M}^{-1})_{11}$, where ${\rm \bf{M}}$ is composed of the elements
\begin{equation}
{\rm \bf{M}}_{ij}=z\delta_{ij}-\langle{i{\rm\bf L}\hat e}_j|{\hat
e}_i\rangle=z\delta_{ij}+\langle{\hat e}_j|i{\rm\bf L}{\hat
e}_i\rangle,  \label{matrixm}
\end{equation}
once a complete basis set, $\{{\hat
e}_\ell|\ell=1,\cdots,\infty\}$, spanning the Liouville space is
given. We select ${\hat e}_1=c_{d\sigma}$ to obtain
$G_{dd\sigma}^+(\omega)$. The inner product in Eq. (\ref{matrixm})
is defined by $\langle\hat e_k|\hat e_\ell\rangle\equiv
\langle\{\hat e_k,\hat e_\ell^\dagger\}\rangle$, where the angular
and curly brackets denote the statistical average and
anticommutator, respectively. We now present a systematic method
for determining the basis vectors. For this purpose, expressing
the Green's function operator in a diagonal matrix as
\begin{equation}
iG_{ij\sigma}^+(z)= (\langle{\hat A}| \, \, \,\langle{\hat B}|)
\left(\begin{array}{cc} {\widehat G}_A \, \, \, \, \, \,  0\, \,
\\ 0 \, \, \, \, \, \, \, \,  {\widehat G}_B
\end{array} \right)\left(\begin{array}{c} |{\hat A}\rangle \\ |{\hat B}\rangle
\end{array} \right) \label{matrixform}
\end{equation}
is the most crucial step. Once this form is achieved, the
Liouville space of $G_{ij\sigma}^+(z)$ will be spanned completely
by the linearly independent components of vectors $|{\hat
A}\rangle$ and $|{\hat B}\rangle$. To obtain the form in Eq.
(\ref{matrixform}), we expand the Green's function operator
${\widehat G}=[z{\bf I}+i({\bf L}_I+{\bf L}_C)]^{-1}$, where ${\bf
L}_I$ and ${\bf L}_C$ respectively represent the Liouville
operators for the isolated part ${\cal H}_I$ and the connecting
part ${\cal H}_C$ of the Hamiltonian; The operator is expanded in
powers of ${\bf L}_C$ using the operator identity $({\hat A}+{\hat
B})^{-1}={\hat A}^{-1}-{\hat A}^{-1}{\hat B}({\hat A}+{\hat
B})^{-1}$, where ${\hat A}=z{\bf I}+i{\bf L}_I$ and ${\hat
B}=i{\bf L}_C$. Then, after resumming the expansion, the retarded
Green's function can be expressed as follows:
\begin{eqnarray}
iG_{ij\sigma}^+(z)&=&\langle
c_{i\sigma}|\widehat{G}_{I}|c_{j\sigma}\rangle -\langle
c_{i\sigma}|\widehat{G}_{I}|i{\rm \bf
L}_{C}\widehat{G}_{I}c_{j\sigma}\rangle\nonumber \\
&+&\langle c_{i\sigma}\widehat{G}_{I}i{\rm \bf
L}_{C}|\widehat{G}|i{\rm \bf
L}_{C}\widehat{G}_{I}c_{j\sigma}\rangle, \label{expand2}
\end{eqnarray}
where ${\widehat G}_I\equiv (z{\bf I}+i{\bf L}_I)^{-1}$. Equation
(\ref{expand2}) can be written in the matrix form as
\begin{equation}
iG_{ij\sigma}^{+}(z)=\left( \langle c_{i\sigma}| \, \, \,
 \langle{\Phi_i}| \right) {\sf G} \left(
|c_{j\sigma}\rangle \, \, \, |{\Phi_j} \rangle \right)^T,
\label{green2}
\end{equation}
where ${\sf G}=\left(
\begin{array}{cc} {\widehat G}_I \, \, \,  -{\widehat G}_I\, \,
\\ 0 \, \,\, \, \, \, \, \, \, \, \, \,   {\widehat G}
\end{array} \right)$,
$|\Phi_{j} \rangle=|i{\bf L}_C{\widehat G}_{I}c_{j\sigma}\rangle$,
 and the superscript $T$ denotes the transpose.
Using the linear transformation ${\sf U}=\left(
\begin{array}{lr} I \, \, \,  -{\widehat G}_{LT}\, \,
\\ 0 \, \,\, \, \, \, \, \, \, \, \, \,   I
\end{array} \right)$, where ${\widehat G}_{LT}={\widehat G}_{I}
/[{\widehat G}-{\widehat G}_{I}]$, one can diagonalize ${\sf G}$.
Equation (\ref{green2}) is then modified as
\begin{equation}
iG_{ij\sigma}^{+}(z)=\left( \langle \widetilde{c}_{i\sigma}| \, \,
\,
 \langle{\Phi_i}| \right) {\sf G}_D \left(
|\widetilde{c}_{j\sigma}\rangle \, \, \, |\Phi_j\rangle \right)^T,
\label{green}
\end{equation}
where ${\sf G}_D=\left(
\begin{array}{cc} {\widehat G}_I \, \, \, \, \, \,  0\, \,
\\ 0 \, \, \, \, \, \, \, \,  {\widehat G}
\end{array} \right)$  and
\begin{eqnarray}
|{\widetilde c}_{j\sigma}\rangle&=&|c_{j\sigma}\rangle+({\widehat
G}_{I}-{\widehat G})^{-1}{\widehat G}_{I}|\Phi_j\rangle \nonumber
\\
&=&|c_{j\sigma}\rangle+|{\rm\bf L}_{C}^{-1}(-iz{\rm\bf I}+{\rm\bf
L})\Phi_j\rangle .\label{newvector}
\end{eqnarray}
Equation (\ref{green}) is the desired form and the linearly
independent components of vectors $|{\widetilde
c}_{j\uparrow}\rangle$ and $|\Phi_{j}\rangle$ completely span the
Liouville space of $G_{ij\sigma}^{+}(\omega)$. In conclusion, the
systematic method for collecting the basis vectors involves
determining all linearly independent components comprising the
vector $|{\widetilde c}_{j\sigma}\rangle$ because $|\Phi_j\rangle$
is contained in this vector.

We will subsequently demonstrate the determination of the basis
vectors by means of an example. We are interested in a mesoscopic
system with a mediating Kondo atom between two metallic
reservoirs. This system can be described by a single-impurity
Anderson model with two metallic reservoirs, whose Hamiltonian is
expressed as
\begin{eqnarray}
{\cal
H}&=&\sum_{k,\sigma,\alpha=L,R}\epsilon_kc^{\alpha\dagger}_{k\sigma}
c^{\alpha}_{k\sigma}+ \sum_{\sigma}\epsilon_dc^\dagger_{d\sigma}
c_{d\sigma}+Un_{d\uparrow}n_{d\downarrow} \nonumber \\
&+&\sum_{k,\sigma,\alpha=L,R}(V_{kd}c^\dagger
_{d\sigma}c^\alpha_{k\sigma}+V^*_{kd}c^{\alpha\dagger}_{k\sigma}
c_{d\sigma}), \nonumber
\end{eqnarray}
where $\epsilon_{k}$, $\epsilon_{d}$, $V_{kd}$, and $U$ indicate
the energies of an electron of momentum $k$ in lead, the level of
a mediating atom, the hybridization between the atom and lead, and
the on-site Coulomb repulsion at the atom, respectively. Because
the two-reservoir Anderson model is a straightforward extension of
the one-reservoir model, we consider a one-reservoir Anderson
model as an example. Its isolated and coupled parts are given by
\begin{equation}
{\cal H}_I=\sum_{k,\sigma}\epsilon_kc^\dagger_{k\sigma}
c_{k\sigma}+\sum_\sigma\epsilon_dc^\dagger _{d\sigma}c_{d\sigma}+
Un_{d\uparrow}n_{d\downarrow}, \label{isol}
\end{equation}
and
\begin{equation}
{\cal H}_C = \sum_{k,\sigma}(V_{kd}c^\dagger
_{d\sigma}c_{k\sigma}+V^*_{kd}c^\dagger_{k\sigma}c_{d\sigma}).
\label{connect}
\end{equation}

We first demonstrate the calculation of $\Phi_d=i{\bf L}_C({\bf
I}+i{\bf L}_I)^{-1}c_{d\uparrow}$ and then of $|{\widetilde
c}_{d\sigma}\rangle$ for ${\cal H}_I$ and ${\cal H}_C$ above. It
can be clearly observed that $({\bf I}+i{\bf
L}_I)^{-1}c_{d\uparrow}$ yields only two operators,
$c_{d\uparrow}$ and $n_{d\downarrow}c_{d\uparrow}$. Applying
${\rm\bf L}_{C}$ to these operators changes the index $d (k)$ into
$k (d)$ in fermion operators and $n_{d\downarrow}$ into
$j^-_{d\downarrow}$. Therefore, the components of $\Phi_d$ are
given by
\begin{equation}
\Phi_d=(c_{k\uparrow}, \hspace{0.3cm}
n_{d\downarrow}c_{k\uparrow}, \hspace{0.3cm}
j^-_{d\downarrow}c_{d\uparrow}), \label{phid}
\end{equation}
where $k=1, 2, \cdots, \infty$ and
\begin{equation}j^-_{d\downarrow}=i\sum_k(V_{kd}c^\dagger
_{d\downarrow}c_{k\downarrow}-V^*_{kd}c^\dagger_{k\downarrow}c_{d\downarrow}).\end{equation}
We now focus on the vector $|{\widetilde c}_{d\sigma}\rangle$ in
Eq. (\ref {newvector}). The operator ${\bf
L}\Phi_d=[H,c_{k\uparrow}]+[H,n_{d\downarrow}c_{k\uparrow}]+[H,j^-_{d\downarrow}c_{d\uparrow}]$
yields $c_{d\uparrow}$ and $c_{k\uparrow}$ from the first
commutator; $n_{d\downarrow}c_{d\uparrow}$,
$n_{d\downarrow}c_{k\uparrow}$, and
$j^-_{d\downarrow}c_{k\uparrow}$ from the second commutator; and
$j^-_{d\downarrow}c_{d\uparrow}$,
$j^-_{d\downarrow}c_{k\uparrow}$,
$j^+_{d\downarrow}c_{d\uparrow}$,
$j^-_{d\downarrow}n_{d\downarrow}c_{d\uparrow}$, and
$j^+_{d\downarrow}n_{d\uparrow}c_{d\uparrow}$ from the third
commutator, where
\begin{equation}j^+_{d\downarrow}=\sum_k(V_{kd}c^\dagger
_{d\downarrow}c_{k\downarrow}+V^*_{kd}c^\dagger_{k\downarrow}c_{d\downarrow}).\end{equation}
The last operator $j^+_{d\downarrow}n_{d\uparrow}c_{d\uparrow}$ is
a vanishing one, and the operator
$j^-_{d\downarrow}n_{d\downarrow}c_{d\uparrow}$ is dynamically
equivalent to $j^-_{d\downarrow}c_{d\uparrow}$. Thus, the linearly
independent components of $({\bf I}+{\bf L})\Phi_d$ are classified
into two groups, one involving $c_{k\uparrow}$ and the other
involving $c_{d\uparrow}$. These components are
\begin{equation} ({\bf I}+{\bf L})\Phi_d^{\rm k}=(c_{k\uparrow}, \hspace{0.3cm}
n_{d\downarrow}c_{k\uparrow}, \hspace{0.3cm}
j^-_{d\downarrow}c_{k\uparrow}), \label{phiLd}
\end{equation}
for $k=1, 2, \cdots, \infty$, and
\begin{equation}
({\bf I}+{\bf L})\Phi_d^{\rm d}=(c_{d\uparrow}, \hspace{0.3cm}
n_{d\downarrow}c_{d\uparrow}, \hspace{0.3cm}
j^-_{d\downarrow}c_{d\uparrow}, \hspace{0.3cm}
j^+_{d\downarrow}c_{d\uparrow}). \label{phiLd2}
\end{equation}

Finally, we apply ${\bf L}_C^{-1}$, which is equivalent to the
repeated application of ${\bf L}_C$, to the operators in Eqs.
(\ref{phiLd}) and (\ref{phiLd2}). It can be clearly observed that
the multiple application of ${\bf L}_C$ to $c_{k\uparrow}$,
$c_{d\uparrow}$, and $n_{d\uparrow}$ simply reproduces the
existing operators and $j^+_{d\downarrow}c_{k\uparrow}$. The
remaining linearly independent operators arise from ${\bf
L}_C^{n}j^\mp_{d\downarrow}$, where $n=1, 2, \cdots, \infty$.
Thus, one may classify the basis vectors into two groups,
\begin{eqnarray}
\mbox{Set of} \, \, c_{d\uparrow}&=&\{c_{d\uparrow},
\hspace{0.3cm} n_{d\downarrow}c_{d\uparrow}, \hspace{0.3cm}
j^-_{d\downarrow}c_{d\uparrow}, \hspace{0.3cm}
j^+_{d\downarrow}c_{d\uparrow}, \hspace{0.3cm} \nonumber \\
& &({\bf L}_C^{n}j^\mp_{d\downarrow})c_{d\uparrow},\hspace{0.3cm}
({\bf L}_C^{n}j^-_{d\downarrow}n_{d\downarrow})c_{d\uparrow}\}
\label{phiLd3}
\end{eqnarray}
and
\begin{eqnarray}
\mbox{Set of} \, \, c_{k\uparrow}&=&\{c_{k\uparrow},
\hspace{0.3cm} n_{d\downarrow}c_{k\uparrow}, \hspace{0.3cm}
j^-_{d\downarrow}c_{k\uparrow}, \hspace{0.3cm}
j^+_{d\downarrow}c_{k\uparrow}, \hspace{0.3cm}\nonumber \\
& & ({\bf L}_C^{n}j^\mp_{d\downarrow})c_{k\uparrow}\}.
\label{phiLd4}
\end{eqnarray}
The operators shown in Eqs. (\ref{phiLd3}) and (\ref{phiLd4})
describe all the possible linearly independent ways of
annihilating an up-spin at site $d$ in time $t$. In other words,
these operators completely span the Liouville space of
$c_{d\uparrow}(t)$.

It is important to identify the meaning of the basis vectors
$({\bf L}_C^{n}j^\mp_{d\downarrow})c_{k\uparrow}$ and $({\bf
L}_C^{n}j^\mp_{d\downarrow})c_{d\uparrow}$ for later discussion.
Application of ${\bf L}_C$ to $j^\mp_{d\downarrow}$ gives rise to
\begin{eqnarray*}
{\bf L}_Cj^-_{d\downarrow}&=&-i\sum_{\bf l}\sum_{\bf k}(V_{{\bf
l}d}V_{{\bf k}d}^* c^\dagger_{{\bf l}\downarrow}c_{{\bf
k}\downarrow}+V_{{\bf k}d}V_{{\bf l}d}^*c^\dagger_{{\bf
k}\downarrow}c_{{\bf l}\downarrow})\\
&+& 2i\sum_{\bf k}|V_{{\bf k}d}|^2
c^\dagger_{d\downarrow}c_{d\downarrow}\end{eqnarray*} and
$${\bf L}_Cj^+_{d\downarrow}=\sum_{\bf l}\sum_{\bf k}V_{{\bf
l}d}V_{{\bf k}d}^* c^\dagger_{{\bf l}\downarrow}c_{{\bf
k}\downarrow}-\sum_{\bf l}\sum_{\bf k}V_{{\bf l}d}^*V_{{\bf k}d}
c^\dagger_{{\bf k}\downarrow}c_{{\bf l}\downarrow}.$$ These
operators represent a round trip of a down-spin electron between
the mediating atom and electron reservoir. Therefore, $({\bf
L}_C^{n}j^\mp_{d\downarrow})$ depicts $n$-time trip of a down-spin
electron between the mediating atom and the reservoir without
coming into contact with an up-spin electron at the mediating
atom. This phenomenon rarely occurs in reality, and we neglect
these basis vectors.

\section{Retarded Green's Function for Two-electrode System}

We identify the basis vectors that do not play an important role
in describing the dynamics $c_{d\uparrow}(t)$ in the Kondo regime.
We first consider the basis vector $n_{d\downarrow}c_{d\uparrow}$
in Eq. (\ref{phiLd3}). Since $n_{d\downarrow}^2=n_{d\downarrow}$,
this basis vector describes the dynamical processes involving all
higher orders of Coulomb interaction $U$. Therefore, in the Kondo
regime, $n_{d\downarrow}c_{d\uparrow}$ must be eliminated. The
operator $j^-_{d\downarrow}n_{d\downarrow}c_{d\uparrow}$ in Eq.
(\ref{phiLd3}) also represents the process that costs energy $U$
or more. Therefore, we eliminate the basis vectors $({\bf
L}_C^{n}j^-_{d\downarrow}n_{d\downarrow})c_{d\uparrow}$ in
studying the Kondo regime. We further eliminate the basis vectors
$({\bf L}_C^{n}j^\mp_{d\downarrow})c_{d\uparrow}$ because multiple
round trip of an up-spin electron without coming into contact with
a down-spin electron at the mediating atom are practically rare.
These approximations are valid for describing the Kondo processes.
Now, we discuss on the basis vectors combined with $c_k$ shown in
Eq. (\ref{phiLd4}). These basis vectors contribute to constructing
the self-energy. It is sufficient to choose $c_k$ and
$n_{d\downarrow}c_{k\uparrow}$ in Eq. (\ref{phiLd4}) to construct
the self-energy. The basis vectors $({\bf
L}_C^{n}j^\mp_{d\downarrow})c_{k\uparrow}$ are eliminated by the
same reason eliminating $({\bf
L}_C^{n}j^\mp_{d\downarrow})c_{d\uparrow}$ in Eq. (\ref{phiLd3})
and the contribution by $j^\mp_{d\downarrow}c_{k\uparrow}$ is much
smaller than that by $n_{d\downarrow}c_{k\uparrow}$.

Finally, we obtain the reduce Liouville space constructed by the
basis vectors
$$(c_{d\uparrow}, \hspace{0.3cm}
j^-_{d\downarrow}c_{d\uparrow}, \hspace{0.3cm}
j^+_{d\downarrow}c_{d\uparrow}) \hspace{0.3cm} \mbox{\rm and}
\hspace{0.3cm} (c_{k\uparrow}, \hspace{0.3cm}
n_{d\downarrow}c_{k\uparrow}),$$ where $k=1, 2, \cdots, \infty$
for describing the dynamics of the single-impurity Anderson model
in the Kondo regime.

Since the resolvent form of $G_{dd\uparrow}^{+}(z)$ is given by
the Laplace transform of the coefficient $a_1(t)$ in the expansion
$c_{d\uparrow}(t)=\sum_{i=1}^\infty a_i(t){\hat e}_i$, where
${\hat e}_1=c_{d\uparrow}$, the equality
$a_1(t)=G_{dd\uparrow}^{+}(t)$ must be obtained. This equality is
valid when all the basis vectors except $c_{d\uparrow}$ are
orthogonal to $c_{d\uparrow}$. This orthogonality condition is
satisfied by introducing the expression $\delta {\hat A}={\hat
A}-\langle {\hat A}\rangle$ for the operators $n_{d\uparrow}$ and
$j^\mp_{d\uparrow}$. Thus, we finally obtain the reduced Liouville
space spanned by the basis vectors
\begin{eqnarray*}(c_{d\uparrow}, \hspace{0.2cm} \delta
j^-_{d\downarrow}c_{d\uparrow}, \hspace{0.2cm} \delta
j^+_{d\downarrow}c_{d\uparrow}) \hspace{0.2cm} \mbox{\rm and}
\hspace{0.2cm} (c_{k\uparrow}, \hspace{0.2cm} \delta
n_{d\downarrow}c_{k\uparrow}), \label{basis}
\end{eqnarray*}
where $k=1, 2, \cdots, \infty$, for the one-reservoir Anderson
model in the Kondo regime. In practical calculations, we use the
normalized forms of the basis vectors. The basis vectors given
above are the same as those used in the previous
study~\cite{hong07}, in which the basis vectors were obtained
intuitively. The Liouville space for the two-reservoir Anderson
model is a straightforward extension of that for the
single-reservoir model. The basis vectors are given by
$$(c_{k\uparrow}^L, \, \, \delta n_{d\downarrow}c_{k\uparrow}^L),
\, \, \mbox{where} \, \,  \, \, k=1, 2, \cdots, \infty; \mbox{and}
$$
$$(\delta j^{+L}_{d\downarrow}c_{d\uparrow}, \, \, \delta
j^{-L}_{d\downarrow}c_{d\uparrow}, \, \, c_{d\uparrow}, \, \,
\delta j^{-R}_{d\downarrow}c_{d\uparrow}, \, \, \delta
j^{+R}_{d\downarrow}c_{d\uparrow})  \, \,  \, \, \mbox{and} $$
$$(\delta n_{d\downarrow}c_{k\uparrow}^R, \, \, c_{k\uparrow}^R),
\, \, \mbox{where} \, \,  \, \, k=1, 2, \cdots, \infty,$$ where
the superscripts $L$ and $R$ denote the left and right metallic
leads, respectively. The matrix ${\rm\bf M}$ of Eq.
(\ref{matrixm}) is expressed as
\begin{eqnarray} {\rm\bf M}=\left(
\begin{array}{ccc} {\rm\bf M}_{LL} & {\rm\bf M}_{dL} &
{\rm\bf 0} \\ {\rm\bf M}_{Ld} & {\rm\bf M}_{d} & {\rm\bf M}_{Rd}
\\ {\rm\bf 0} & {\rm\bf M}_{dR}
&  {\rm\bf M}_{RR}
\end{array} \right),
\end{eqnarray}
where ${\rm\bf M}_{d}$ is the $5\times 5$ block that is
constructed by the basis vectors
$$ (\delta j^{+L}_{d\downarrow}c_{d\uparrow}, \hspace{0.3cm}
\delta j^{-L}_{d\downarrow}c_{d\uparrow}, \hspace{0.3cm}
c_{d\uparrow}, \hspace{0.3cm} \delta
j^{-R}_{d\downarrow}c_{d\uparrow}, \hspace{0.3cm} \delta
j^{+R}_{d\downarrow}c_{d\uparrow}); $$ ${\rm\bf M}_{LL}$, the
$\infty\times\infty$ block constructed by the basis vectors
$(c_{k\uparrow}^L, \, \, \delta n_{d\downarrow}c_{k\uparrow}^L),$
where $k=1, 2, \cdots, \infty$; and ${\rm\bf M}_{dL}$, for
example, a $5\times\infty$ block constructed  by the basis vectors
given above.

The retarded Green's function $G_{dd\uparrow}^{+}(\omega)$ is
obtained by calculating the matrix inverse $({\rm\bf
M}^{-1})_{dd}$, where the subscript $dd$ indicates the element
obtained from $c_{d\uparrow}$ in both row and column. Calculating
the inverse of an infinite-dimensional matrix ${\rm\bf M}$ is a
nontrivial problem. For this purpose, we perform matrix reduction
by using L\"owdin's partitioning technique~\cite{loedin,mujika}.
We consider an eigenvalue equation ${\rm\bf M}{\rm\bf C}={\rm\bf
0}$, in which the column vector ${\rm\bf C}=({\rm\bf C}_L \,
{\rm\bf C}_d \, {\rm\bf C}_R)^T$, where ${\rm\bf C}_L$, ${\rm\bf
C}_d $, and ${\rm\bf C}_R$ correspond to the row vectors
$(c_{k\uparrow}^L, \delta n_{d\downarrow}c_{k\uparrow}^L)$,
$(\delta j^{+L}_{d\downarrow}c_{d\uparrow}, \delta
j^{-L}_{d\downarrow}c_{d\uparrow}, c_{d\uparrow}, \delta
j^{-R}_{d\downarrow}c_{d\uparrow}, \delta
j^{+R}_{d\downarrow}c_{d\uparrow})$, and $(\delta
n_{d\downarrow}c_{k\uparrow}^R, c_{k\uparrow}^R)$, respectively,
and ${\rm\bf 0}$ is a column vector with zero elements. After
eliminating ${\rm\bf C}_{L}$ and ${\rm\bf C}_{R}$, we obtain an
equation $({\rm\bf M}_{d}-{\rm\bf M}_{Ld}{\rm\bf
M}_{LL}^{-1}{\rm\bf M}_{dL}-{\rm\bf M}_{Rd}{\rm\bf
M}_{RR}^{-1}{\rm\bf M}_{dR}){\rm\bf C}_{d}\equiv {\rm\bf
M}_{r}{\rm\bf C}_{d}={\rm\bf 0}$, which leads to
\begin{equation}
{\rm\bf M}_{r}={\rm\bf M}_{d}-{\rm\bf M}_{Ld}{\rm\bf
M}_{LL}^{-1}{\rm\bf M}_{dL}-{\rm\bf M}_{Rd}{\rm\bf
M}_{RR}^{-1}{\rm\bf M}_{dR}.\label{eigen22}
\end{equation}
The last two terms in Eq. (\ref{eigen22}) contribute to the
self-energy of the retarded Green's function. We finally obtain
${\rm\bf M}_{r}$ for the two-reservoir Anderson model as follows:
\begin{eqnarray} {\rm\bf M}_{r}
=\left( \begin{array}{c l l c c} -i\omega' & -\gamma_{LL} &
U^L_{J^-} & \gamma_{LR} & \gamma_{J^-} \\ \gamma_{LL}  & -i\omega'
&
 U^L_{J^+} & \gamma_{J^+} & \gamma_{LR} \\
-U_{J^-}^{L*} &  -U_{J^+}^{L*} & -i\omega' &  -U^{R*}_{J^+} &
-U^{R*}_{J^-} \\  -\gamma_{LR} & -\gamma_{J^-} & U_{J^+}^R  &
-i\omega' & \gamma_{RR} \\
 -\gamma_{J^+} &  -\gamma_{LR} &
 U_{J^-}^R  & -\gamma_{RR} &  -i\omega'
\end{array} \right). \label{r2r}
\end{eqnarray}
Here, $\omega'\equiv\omega-\epsilon_d-U\langle
n_{d\downarrow}\rangle$, where $\epsilon_{d}$, $U$, and $\langle
n_{d\downarrow}\rangle$ denote the energy level of the Kondo
impurity, Coulomb interaction, and the average number of down-spin
electrons occupying the level $\epsilon_{d}$, respectively. The
derivations of matrix elements are given in Appendix A. All the
matrix elements, except the eight $U$-elements, have additional
self-energy terms
$i\Sigma_{mn}=\beta_{mn}(i\Sigma^L_0(\omega)+i\Sigma^R_0(\omega))=2\beta_{mn}\Delta$,
where $\Sigma_0^L(\omega)=\sum_{\bf
k}|V_{kd}^L|^2/(\omega-\epsilon_{\bf k}+i\eta)=-i\Delta^L $ for a
flat wide-band and $\Delta\equiv (\Delta^L+\Delta^R)/2$. $\Delta$
is used as a unit of energy. The lead function $\Gamma^L(\omega)$
in Eq. (1) is equal to 2$\Delta^L$ in this study. The coefficients
$\beta_{mn}$ are $\beta_{11}=|\xi_d^{-L}|^2$,
$\beta_{22}=|\xi_d^{+L}|^2$, $\beta_{33}=1$,
$\beta_{44}=|\xi_d^{+R}|^2$, $\beta_{55}=|\xi_d^{-R}|^2$,
$\beta_{12(21)}=[\xi_d^{-L*}\xi_d^{+L}]^{(*)}$,
$\beta_{14(41)}=[\xi_d^{-L*}\xi_d^{+R}]^{(*)}$,
$\beta_{15(51)}=[\xi_d^{-L*}\xi_d^{-R}]^{(*)}$,
$\beta_{24(42)}=[\xi_d^{+L*}\xi_d^{+R}]^{(*)}$,
$\beta_{25(52)}=[\xi_d^{+L*}\xi_d^{-R}]^{(*)}$, and
$\beta_{45(54)}=[\xi_d^{+R*}\xi_d^{-R}]^{(*)}$, where $\xi_d^{\mp
L}$, for example, is expressed as
\begin{eqnarray}\xi_d^{\mp
L}&=&\frac{1}{2}\frac{\langle i[n_{d\downarrow},j^{\mp
L}_{d\downarrow}] (1-2n_{d\uparrow})\rangle+i(1-2\langle
n_{d\downarrow}\rangle)\langle j^{\mp
L}_{d\downarrow}\rangle}{\langle (\delta j^{\mp
L}_{d\downarrow})^2\rangle^{1/2}\langle (\delta
n_{d\downarrow})^2\rangle^{1/2}} \nonumber \\ &=&
U_{J^\mp}^{L}/[U\langle (\delta
n_{d\downarrow})^2\rangle^{1/2}].\label{22}
\end{eqnarray}
We show the explicit expressions of $\beta_{mn}$ in Appendix B.

The matrix elements represented by $\gamma$ are written as
\begin{eqnarray}\gamma_{LL(RR)}&=&\langle \sum_k
i(V_{kd}^{*}c_{k\uparrow}^{L}+V_{kd}^{*}c_{k\uparrow}^{R})
c^\dagger_{d\uparrow}[j^{-L(R)}_{d\downarrow},j^{+L(R)}_{d\downarrow}]
\rangle \nonumber \\
&\times& [\langle (\delta j^{-L(R)}_{d\downarrow})^2\rangle\langle
(\delta
j^{+L(R)}_{d\downarrow})^2\rangle]^{-1/2},\label{23}\end{eqnarray}
\begin{eqnarray}\gamma_{J^\mp}&=&\langle\sum_ki
(V_{kd}^{*}c_{k\uparrow}^L+V_{kd}^{*}c_{k\uparrow}^R)
c^\dagger_{d\uparrow}[j^{\mp L}_{d\downarrow},j^{\mp
R}_{d\downarrow}]\rangle \nonumber \\
&\times& [\langle (\delta j^{\mp L}_{d\downarrow})^2\rangle\langle
(\delta j^{\mp
R}_{d\downarrow})^2\rangle]^{-1/2},\label{24}\end{eqnarray} and
\begin{eqnarray}\gamma_{LR}&=&\langle\sum_ki
(V_{kd}^{*}c_{k\uparrow}^L+V_{kd}^{*}c_{k\uparrow}^R)c^\dagger_{d\uparrow}
[j^{-L}_{d\downarrow},j^{+R}_{d\downarrow}]\rangle \nonumber \\
&\times& [\langle (\delta j^{-L}_{d\downarrow})^2\rangle\langle
(\delta
j^{+R}_{d\downarrow})^2\rangle]^{-1/2}.\label{25}\end{eqnarray} We
show the detailed derivations in Appendix A. The retarded Green's
function $iG_{dd\uparrow}^{+}(\omega)=({\rm\bf M}_r^{-1})_{33}$
may explain all the features of nonlinear conductance observed in
a quantum point contact\cite{cronen,dicarlo,sfigakis,sarkozy}, a
single electron transistor\cite{cronen2,kogan,amasha,nygard}, and
scanning tunneling microscopy on a magnetic atom adsorbed on a
metallic substrate\cite{mad,mano,neel,otte}.

\begin{figure}
[t] \vspace*{5cm} \includegraphics{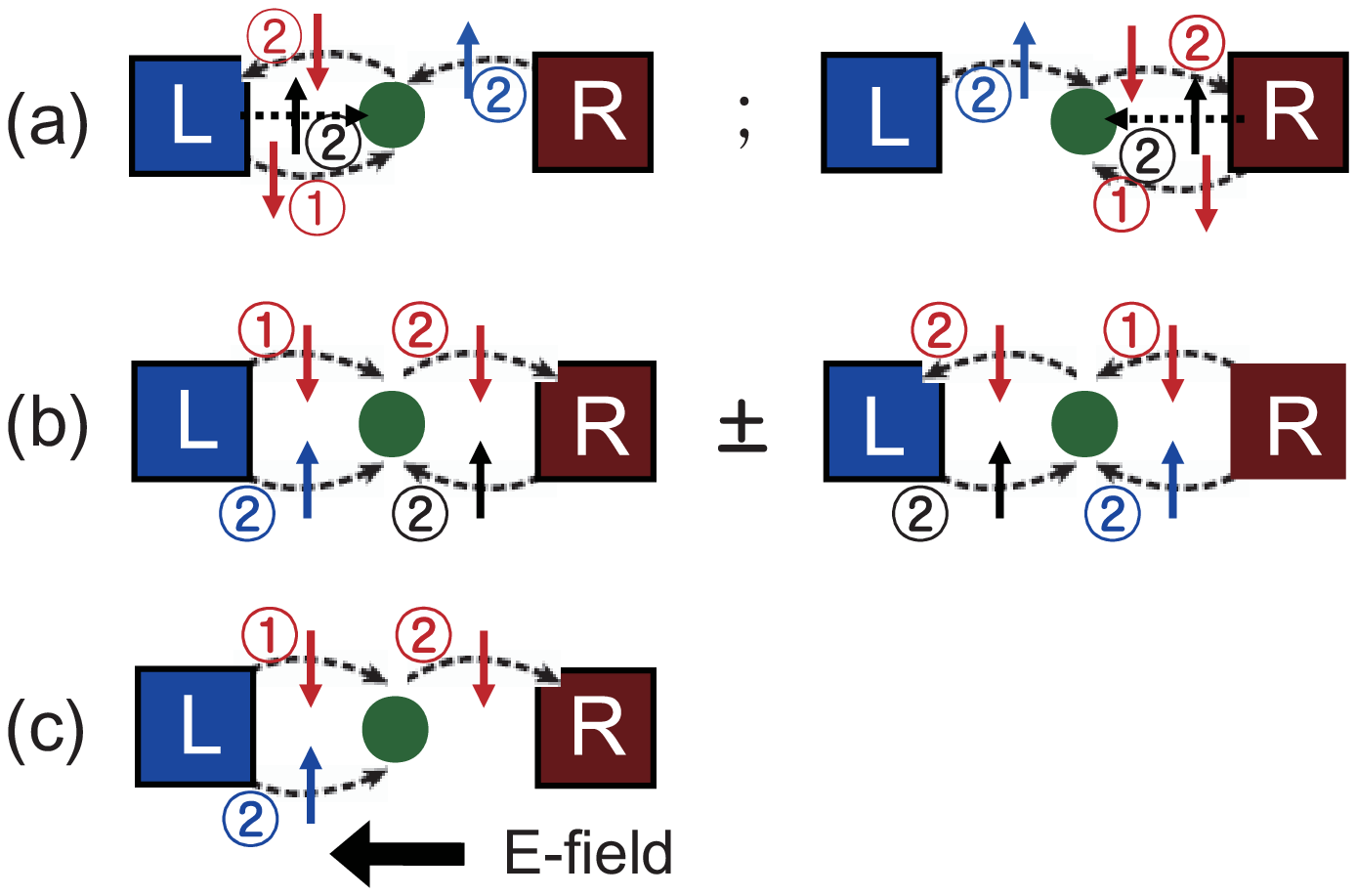} \vspace*{0.0cm}
\caption{Motions of electrons indicated in $\gamma$. (a) Kondo
processes described by $\gamma_{LL}$ (left) and $\gamma_{RR}$
(right). The solid dot indicates a Kondo impurity. The up- and
down-spin motions denoted by $\textcircled2$ on the same side
represent the exchange process while the up- and down-spin motions
denoted by $\textcircled2$ on different sides indicate singlet
hopping. (b) Kondo processes of connecting mechanism:
$\gamma_{LR}$ $(+$ sign) and $\gamma_{J}$ $(-$ sign). (c)
Unidirectional resonant tunneling of a singlet establishing
current flow. }
\end{figure}

In Fig.~1, we present the graphical illustrations of $\gamma$ on
the basis of the corresponding operator expressions, which are
third-order processes of hybridization. Figure~1 depicts the Kondo
processes in which a down-spin indicated by $\textcircled1$ first
enters the Kondo impurity and forms a singlet with an un-spin
electron indicated by $\textcircled2$. Then, it performs an
exchange process or singlet hopping. Figure~1 (a) describes the
spin movements in $\gamma_{LL}$ (left) and $\gamma_{RR}$ (right)
in equilibrium, and Fig.~1 (b), those in $\gamma_{LR}$ $(+)$ and
$\gamma_{J}$ $(-)$. Each $\gamma$ is composed of two terms, i.e.,
spin exchange and singlet hopping. The latter is a unique property
of a two-reservoir system. It is noteworthy that $\gamma_{LL}$ and
$\gamma_{RR}$ represent the degree of Kondo coupling, whereas
$\gamma_{LR}$ and $\gamma_{J}$ represent the resonant tunneling
between two reservoirs.

The effect of bias is contained in the Fermi distribution function
and is observed in the calculation of the expectation values of
$\gamma$. In the ground state, resonant tunneling from the
reservoir at lower chemical potential to the one at higher
chemical potential is prohibited. Therefore, the second part of
Fig.~1 (b) vanishes. The motion that establishes current flow in
both $\gamma_{LR}$ and $\gamma_{J}$ is unidirectional, as shown in
Fig.~1 (c), which leads to the condition $\gamma_{LR}=\gamma_{J}$.
However, when the second part of Fig.~1 (b) does not vanish for
certain reason, the condition $\gamma_{LR}>\gamma_{J}$ is
obtained. In both conditions of $\gamma$, even the state of very
low bias is quite different from that of zero bias at which
$\gamma_{J}=0$.

\section{Self-consistent Method}

The values of $\gamma$ and the fluctuations in the denominators of
Eqs. (\ref{22})$-$(\ref{25}) may not be determined by direct
calculation. These values can be determined by employing the
self-consistent method, which was used by Nagaoka\cite{nagaoka}
for studying the conventional Kondo problem. However, we do not
try to determine these values in this study. Nevertheless, in the
following study, we consider them as free parameters in order to
explain the experimental observations for a quantum point
contact\cite{cronen,dicarlo,sfigakis,sarkozy} and scanning
tunneling microscopy\cite{mad,mano,neel,otte}.

Finally, we suggest a self-consistent scheme that is particulary
useful when the application of a magnetic field disturbs the
particle-hole symmetry. Because the matrix elements
$U^{L,R}_{J^\mp}$ and coefficients $\beta_{mn}$ contain
undetermined quantities $\langle n_{d\sigma}\rangle, \langle
j^{-L,R}_{d\sigma}\rangle$, and $\langle
j^{+L,R}_{d\sigma}\rangle$, and these quantities are also
expressed by the retarded Green's function such as
\begin{eqnarray}
\langle n_{d\sigma}\rangle&=&-\frac{1}{\pi}\int_{-\infty}^\infty
\frac{f_L(\omega)\Gamma^L(\omega)+ f_R(\omega)\Gamma^R(\omega)}
{\Gamma^L(\omega)+\Gamma^R(\omega)}\nonumber \\
&\times& {\rm Im}\,G_{dd\sigma}^{+}(\omega)d\omega; \, \,
\label{nave}
\end{eqnarray}
\begin{eqnarray}
\langle j^{-L}_{d\sigma}\rangle
&=&-\frac{1}{\pi}\int\frac{d\omega}{2}[f_L(\omega)-f_R(\omega)]\widetilde{\Gamma}(\omega)
\nonumber \\
&\times& {\rm Im}\,G_{dd\sigma}^{+}(\omega) =-\langle
j^{-R}_{d\sigma}\rangle; \label{current2}
\end{eqnarray}
\begin{eqnarray}
\langle j^{+L(R)}_{d\sigma}\rangle=
\int_\infty^\infty\frac{d\omega}{\pi}f_{L(R)}(\omega)\Gamma^{L(R)}(\omega)
{\rm Re}\,{G}_{dd\sigma}^+(\omega), \label{jplus}
\end{eqnarray}
where $\widetilde{\Gamma}(\omega)=\Gamma^L(\omega)
\Gamma^R(\omega)/[\Gamma^L(\omega)+\Gamma^R (\omega)]$, one can
construct a self-consistent loop to obtain the retarded Green's
function as follows:\vspace{0.3cm}

\noindent  $\langle n_{d\downarrow}\rangle^{(0)}, \langle j^{\mp
L,R}_{d\downarrow}\rangle^{(0)} \rightarrow
{G}_{dd\uparrow}^{+(0)}(\omega) \rightarrow
\rho^{(0)}_\uparrow(\omega) \rightarrow \langle
n_{d\uparrow}\rangle^{(0)}, \langle j^{\mp
L,R}_{d\uparrow}\rangle^{(0)} \rightarrow \indent
{G}_{dd\downarrow}^{+(0)}(\omega) \rightarrow
\rho^{(0)}_\downarrow(\omega) \rightarrow \langle
n_{d\downarrow}\rangle^{(1)}, \langle j^{\mp
L,R}_{d\downarrow}\rangle^{(1)} \rightarrow
{G}_{dd\uparrow}^{+(1)}(\omega) \rightarrow
\rho^{(1)}_\uparrow(\omega) \rightarrow \cdots$ \vspace{0.3cm}

\noindent The expressions in Eqs. (\ref{nave}) and
(\ref{current2}) are valid only when
$\Gamma^L(\omega)\propto\Gamma^R(\omega)$. We report the results
for the magnetic-field-induced peak splitting of the differential
conductance in a single-electron transistor with a quantum
dot\cite{cronen2,kogan,amasha,nygard} in terms of this
self-consistent scheme in a separate study.

\section{Conclusions}
In conclusion, we have presented a systematic methodology for
determining basis vectors spanning the Liouville space, which is
the most crucial step in calculating the retarded Green's function
using the resolvent operator; further, we have suggested a
procedure for calculating the retarded Green's function. The
operator method presented in this study has several advantages:
(i) A complete set of basis vectors cab be determined
systematically, and (ii) the physical meanings of the basis
vectors represented by the operators are apparent, allowing the
identification and removal of the unimportant basis vectors for a
particular parameter regime and subsequent construction of a
reduced Liouville space. The on-site retarded Green's function for
the single-impurity Anderson model with two metallic reservoirs is
expressed as an inverse of a $5\times 5$ matrix. We discussed the
characteristics of the matrix elements under the application of a
bias. Finally, we suggest a self-consistent calculation method
that is useful when the system under consideration does not show
particle-hole symmetry. This operator formulation for the retarded
Green's function may be appropriate for studying the dynamics of a
many-body system under steady-state nonequilibrium. In a separate
study, we obtain the spectral function and differential
conductance for a specific system that has a Kondo impurity
between two metallic reservoirs under bias.

\begin{acknowledgments}
The author thanks J Yi and S H Yoon for their valuable
discussions. This work was supported by the Korea Research
Foundation Grant provided by the Korean Government
(KRF-2007-614-C00005).
\end{acknowledgments}

\section*{Appendix A: Calculation of matrix elements}

Calculation of the matrix elements of Eq. (19) for the
single-impurity Anderson model with one reservoir in which
$\gamma_{LR}$ and $\gamma_J^\mp$ do not appear is shown below. One
can obtain the expressions of $\gamma_{LR}$ and $\gamma_J^\mp$ via
the same manner shown here. The block ${\bf M}_{LL(RR)}$ comprises
two diagonal blocks whose elements are the same and given by
$-i\omega+i\epsilon_1+0^+$, $-i\omega+i\epsilon_2+0^+$, and so on.
Since the calculation
is very simple, we skip it.\\
\noindent (1) Matrix elements of the block ${\bf M}_{Ld}$:\\
\noindent Nontrivial elements of ${\bf M}_{Ld}$ are \\ $-\langle
\{i{\bf L}(c_{k\uparrow}\delta n_{d\downarrow}), \delta
j^\mp_{d\downarrow}c_{d\uparrow}^\dagger\}\rangle$, i.e.,
$\langle\{i[H,c_{k\uparrow}\delta
n_{d\downarrow}],j^\mp_{d\downarrow}c_{d\uparrow}^\dagger\}\rangle\\
=\langle\{i[H,c_{k\uparrow}]\delta
n_{d\downarrow}+c_{k\uparrow}i[H,\delta n_{d\downarrow}],\delta
 j^\mp_{d\downarrow}c_{d\uparrow}^\dagger\}\rangle \\
=-i\epsilon_k\langle\{c_{k\uparrow}\delta n_{d\downarrow}, \delta
j^\mp_{d\downarrow}c_{d\uparrow}^\dagger\}\rangle
-iV_{kd}\langle\{c_{d\uparrow}\delta n_{d\downarrow}, \delta
j^\mp_{d\downarrow}c_{d\uparrow}^\dagger\}\rangle
+\langle\{c_{k\uparrow}\delta j^\mp_{d\downarrow}, \delta
j^\mp_{d\downarrow}c_{d\uparrow}^\dagger\}\rangle$. The first
term\\ $(-i\epsilon_k)\langle\{c_{k\uparrow}\delta
n_{d\downarrow}, \delta
j^\mp_{d\downarrow}c_{d\uparrow}^\dagger\}\rangle
=(-i\epsilon_k)\langle[n_{d\downarrow},j^\mp_{d\downarrow}]
c_{k\uparrow}c_{d\uparrow}^\dagger\rangle$ should vanish because
the up-spin dynamics of this form is not allowed. The third term
also vanishes because $\langle\{c_{k\uparrow}j^\mp_{d\downarrow},
\delta j^\mp_{d\downarrow}c_{d\uparrow}^\dagger\}\rangle =\langle
j^\mp_{d\downarrow}\delta j^\mp_{d\downarrow}
\{c_{k\uparrow},c_{d\uparrow}^\dagger\}\rangle=0$. The second
term, however, must be calculated rigorously, since it is a
hybridization term that is related to the Kondo process. The
operator $\{c_{d\uparrow}\delta n_{d\downarrow}, \delta
j^\mp_{d\downarrow}c_{d\uparrow}^\dagger\}$ is expanded as
$c_{d\uparrow}[\delta n_{d\downarrow}, \delta
j^\mp_{d\downarrow}c_{d\uparrow}^\dagger]+[\delta
j^\mp_{d\downarrow}c_{d\uparrow}^\dagger , c_{d\uparrow}]\delta
n_{d\downarrow}+2c_{d\uparrow}\delta j^\mp_{d\downarrow}
c_{d\uparrow}^\dagger\delta n_{d\downarrow}$. Each term is
rewritten as $c_{d\uparrow}[\delta n_{d\downarrow}, \delta
j^\mp_{d\downarrow}c_{d\uparrow}^\dagger] =
c_{d\uparrow}[n_{d\downarrow},j^\mp_{d\downarrow}]
c_{d\uparrow}^\dagger
=c_{d\uparrow}c_{d\uparrow}^\dagger[n_{d\downarrow},j^\mp_{d\downarrow}]
=(1-n_{d\uparrow})[n_{d\downarrow},j^\mp_{d\downarrow}]$ and
$[\delta j^\mp_{d\downarrow}c_{d\uparrow}^\dagger
,c_{d\uparrow}]\delta n_{d\downarrow}=\delta
j^\mp_{d\downarrow}[c_{d\uparrow}^\dagger ,c_{d\uparrow}] \delta
n_{d\downarrow}=\delta
j^\mp_{d\downarrow}(1-2c_{d\uparrow}c_{d\uparrow}^\dagger)\\
\delta n_{d\downarrow}.$ The last expression
$-2c_{d\uparrow}\delta
j^\mp_{d\downarrow}c_{d\uparrow}^\dagger\delta n_{d\downarrow}$
cancels the third term above. Hence, $-\langle\{i{\bf
L}(c_{k\uparrow}\delta
n_{d\downarrow}),j^\mp_{d\downarrow}c_{d\uparrow}^\dagger\}\rangle
=iV_{kd}\langle(1-n_{d\uparrow})[n_{d\downarrow}, \delta
j^\mp_{d\downarrow}]+\delta j^\mp_{d\downarrow} \delta
n_{d\downarrow}\rangle
=iV_{kd}\langle\left(\frac{1}{2}-n_{d\uparrow}\right)
[n_{d\downarrow}, j^\mp_{d\downarrow}]
+\frac{1}{2}[n_{d\downarrow},j^\mp_{d\downarrow}] +\delta
j^\mp_{d\downarrow}\delta n_{d\downarrow}\rangle
=\frac{iV_{kd}}{2}\langle(1-2n_{d\uparrow})[n_{d\downarrow},
j^\mp_{d\downarrow}]+\{\delta n_{d\downarrow}, \delta
j^\mp_{d\downarrow}\}\rangle=\frac{iV_{kd}}{2}\{\langle
(1-2n_{d\uparrow})[n_{d\downarrow},
j^\mp_{d\downarrow}]\rangle+(1-2\langle
n_{d\downarrow}\rangle)\langle j^\mp_{d\downarrow}\rangle\} .$\\
Therefore, the matrix elements\\ $-\langle\{i{\bf
L}(c_{k\uparrow}\delta n_{d\downarrow}),\delta
j^\mp_{d\downarrow}c_{d\uparrow}^\dagger\}
\rangle/[||c_{k\uparrow}\delta n_{d\downarrow}||\times
||c_{d\uparrow}\delta j^\mp_{d\downarrow}||]=V_{kd}\xi_d^\mp,$
where $\xi_d^\mp=\frac{1}{2}\{\langle
i(1-2n_{d\uparrow})[n_{d\downarrow},j^\mp_{d\downarrow}]\rangle
+i(1-2\langle n_{d\downarrow}\rangle)\langle
j^\mp_{d\downarrow}\rangle\}\langle(\delta
n_{d\downarrow})^2\rangle^{-1/2}\langle( \delta
j^\mp_{d\downarrow})^2\rangle^{-1/2}.$ \vspace{0.5cm}

\noindent  (2) Matrix elements of the block ${\bf M}_{d}$:\\
\noindent (2-1) Diagonal elements: \\
By using the commutator expression $i[H,c_{d\uparrow}]=-i\sum_{\bf
k}V^*_{{\bf k}d} c_{{\bf k}\uparrow}-i\epsilon_d c_{d\uparrow}-iU
c_{d\uparrow} n_{d\downarrow}$, the diagonal elements of the block
${\bf M}_{d}$ are given by\\
\noindent  (A): $-\langle\{i[H,c_{d\uparrow}],
c^\dagger_{d\uparrow}\}\rangle =i\sum_{\bf k}V^*_{{\bf
k}d}\langle\{ c_{{\bf k}\uparrow}, c^\dagger_{d\uparrow}\}\rangle
+i\epsilon_d\langle\{
 c_{d\uparrow}, c^\dagger_{d\uparrow}\}\rangle +iU\langle\{c_{d\uparrow} n_{d\downarrow},
 c^\dagger_{d\uparrow}\}\rangle \\
=i\epsilon_d+iU\langle\{c_{d\uparrow}n_{d\downarrow},
c^\dagger_{d\uparrow}\}\rangle=i\epsilon_d+iU\langle\{c_{d\uparrow},
c^\dagger_{d\uparrow}\} n_{d\downarrow}\rangle \\
=i\epsilon_d+iU\langle n_{d\downarrow}\rangle$, and\\
\noindent (B): $-\langle\{i[H,c_{d\uparrow}\delta
j^\mp_{d\downarrow}], (c_{d\uparrow}\delta
j^\mp_{d\downarrow})^\dagger\}\rangle \\
=-\langle\{i[H,c_{d\uparrow}]\delta j^\mp_{d\downarrow}, \delta
j^\mp_{d\downarrow}c_{d\uparrow}^\dagger\}\rangle-
\langle\{c_{d\uparrow}i[H,\delta j^\mp_{d\downarrow}], \delta
j^\mp_{d\downarrow}c_{d\uparrow}^\dagger\}\rangle$.
The first term of (B) is rewritten as \\
$\langle\{i[H,c_{d\uparrow}]\delta j^\mp_{d\downarrow}, \delta
j^\mp_{d\downarrow}c_{d\uparrow}^\dagger\}\rangle
=\langle\{(-i\sum_{\bf k}V^*_{{\bf k}d}c_{{\bf k}\uparrow}
-i\epsilon_d c_{d\uparrow}-iUc_{d\uparrow}n_{d\downarrow}) \delta
j^\mp_{d\downarrow},(c_{d\uparrow}\delta
j^\mp_{d\downarrow})^\dagger\}\rangle
=-i\epsilon_d\langle\{c_{d\uparrow}\delta j^\mp_{d\downarrow},
\delta j^\mp_{d\downarrow}c_{d\uparrow}^\dagger\}\rangle
-iU\langle\{c_{d\uparrow}n_{d\downarrow}\delta
j^\mp_{d\downarrow},\delta
j^\mp_{d\downarrow}c_{d\uparrow}^\dagger\}\rangle$. Applying the
decoupling approximation, $n_{d\downarrow}\delta
j^\mp_{d\downarrow}=\langle n_{d\downarrow}\rangle\delta
j^\mp_{d\downarrow}$, to the $U$-term above gives rise to a form
of squared norm for the first term of (B), i.e.,
$\langle\{i[H,c_{d\uparrow}]\delta j^\mp_{d\downarrow}, \delta
j^\mp_{d\downarrow}c_{d\uparrow}^\dagger\}\rangle
=[-i\epsilon_d-iU\langle n_{d\downarrow}\rangle]\times
||c_{d\uparrow}\delta j^-_{d\downarrow}||^2.$ The second term of
(B), however, cannot be written in the form of a squared norm,
because $[H,j^\mp_{d\downarrow}]\propto j^\pm_{d\downarrow}$.
Therefore, we neglect it.

\noindent (2-2) Matrix elements $U_{J^\mp}$:\\
These are given by the inner products \\ $-\langle\{ i{\bf
L}c_{d\uparrow}, \delta
j^\mp_{d\downarrow}c_{d\uparrow}^\dagger\}\rangle$, i.e.,
$-\langle\{ i[H,c_{d\uparrow}], \delta
j^\mp_{d\downarrow}c_{d\uparrow}^\dagger\}\rangle\\
=i\epsilon_d\langle\delta j^\mp_{d\downarrow}
\rangle\langle\{c_{d\uparrow},c_{d\uparrow}^\dagger\}\rangle
+iU\langle\{c_{d\uparrow}n_{d\downarrow},\delta
j^\mp_{d\downarrow} c_{d\uparrow}^\dagger\}\rangle \\+i\sum_{\bf
k}V^*_{{\bf k}d}\langle\delta j^\mp_{d\downarrow}
\rangle\langle\{c_{{\bf
k}\uparrow},c_{d\uparrow}^\dagger\}\rangle
=iU\langle\{c_{d\uparrow}n_{d\downarrow},\delta
j^\mp_{d\downarrow} c_{d\uparrow}^\dagger\}\rangle$. However,
$\langle\{ c_{d\uparrow}\delta n_{d\downarrow},\delta
j^\mp_{d\downarrow}c_{d\uparrow}^\dagger\}\rangle=\langle
c_{d\uparrow}[\delta n_{d\downarrow},\delta
j^\mp_{d\downarrow}c_{d\uparrow}^\dagger]\rangle+\langle[\delta
j^\mp_{d\downarrow}c_{d\uparrow}^\dagger,c_{d\uparrow}] \delta
n_{d\downarrow}\rangle+\langle 2c_{d\uparrow}\delta
j^\mp_{d\downarrow}c_{d\uparrow}^\dagger\delta
n_{d\downarrow}\rangle$, where the operator in the first term is
rewritten as $c_{d\uparrow}[\delta n_{d\downarrow},\delta
j^\mp_{d\downarrow} c_{d\uparrow}^\dagger]
=c_{d\uparrow}[n_{d\downarrow},j^\mp_{d\downarrow}]c_{d\uparrow}^\dagger
=c_{d\uparrow}c_{d\uparrow}^\dagger[n_{d\downarrow},j^\mp_{d\downarrow}]
=(1-n_{d\uparrow})[n_{d\downarrow},j^\mp_{d\downarrow}]$, while
the second term is given by $\langle[\delta
j^\mp_{d\downarrow}c_{d\uparrow}^\dagger ,c_{d\uparrow}] \delta
n_{d\downarrow}\rangle=\langle\delta
j^\mp_{d\downarrow}[c_{d\uparrow}^\dagger ,c_{d\uparrow}]\delta
n_{d\downarrow}\rangle=\langle\delta j^\mp_{d\downarrow}\delta
n_{d\downarrow}\rangle-\langle 2\delta
j^\mp_{d\downarrow}c_{d\uparrow}c_{d\uparrow}^\dagger\delta
n_{d\downarrow}\rangle$. The last one cancels the third term of
$\langle\{ c_{d\uparrow}\delta n_{d\downarrow},\delta
j^\mp_{d\downarrow}c_{d\uparrow}^\dagger\}\rangle$ above. Hence,
$\langle\{c_{d\uparrow}\delta n_{d\downarrow}, \delta
j^\mp_{d\downarrow}c_{d\uparrow}^\dagger\}\rangle
=\langle(1-n_{d\uparrow})[n_{d\downarrow},j^\mp_{d\downarrow}]
+\delta j^\mp_{d\downarrow}\delta n_{d\downarrow}\rangle
=\langle\left(\frac{1}{2}-n_{d\uparrow}\right)[n_{d\downarrow},j^\mp_{d\downarrow}]+\frac{1}{2}[\delta
n_{d\downarrow}, \delta j^\mp_{d\downarrow}]+\delta
j^\mp_{d\downarrow}\delta n_{d\downarrow}\rangle
=\frac{1}{2}\langle(1-2n_{d\uparrow})[n_{d\downarrow},j^\mp_{d\downarrow}]+\{\delta
n_{d\downarrow},\delta j^\mp_{d\downarrow}\}\rangle
=\frac{1}{2}\{\langle(1-2n_{d\uparrow})[n_{d\downarrow},j^\mp_{d\downarrow}]\rangle+(1-2\langle
n_{d\downarrow}\rangle)\langle j^\mp_{d\downarrow}\rangle\}$.
Therefore, the matrix elements denoted by $U_{J^\mp}$ is given by
\begin{eqnarray*}
U_{J^\mp}&=&-\langle \{i{\bf L}c_{d\uparrow},\delta
j^\mp_{d\downarrow}c_{d\uparrow}^\dagger\}\rangle\langle(\delta
j^\mp_{d\downarrow})^2\rangle^{-1/2}
\\
&=&\frac{iU}{2}[\langle(1-2n_{d\uparrow})[n_{d\downarrow},j^\mp_{d\downarrow}]\rangle
\\ &+&(1-2\langle n_{d\downarrow}\rangle)\langle
j^\mp_{d\downarrow}\rangle] \langle(\delta
j^\mp_{d\downarrow})^2\rangle^{-1/2}.\end{eqnarray*}

\noindent (2-3) Matrix elements $\gamma_{LL(RR)}$:\\
The inner products $-\langle\{ i{\bf L}(c_{d\uparrow}\delta
j^\mp_{d\downarrow}), \delta
j^\pm_{d\downarrow}c_{d\uparrow}^\dagger\}\rangle$ gives rise
\\to the expressions for $\gamma_{LL(RR)}$. Since \\ $\langle\{i{\bf
L}(c_{d\uparrow}\delta j^\mp_{d\downarrow}),\delta
j^\pm_{d\downarrow}c_{d\uparrow}^\dagger\} \rangle
=\langle\{i[H,c_{d\uparrow}]\delta j^\mp_{d\downarrow},\delta
j^\pm_{d\downarrow}c_{d\uparrow}^\dagger\}\rangle+\langle\{c_{d\uparrow}i[H,
j^\mp_{d\downarrow}], \delta
j^\pm_{d\downarrow}c_{d\uparrow}^\dagger\}\rangle$, the first term
can be expanded as $\langle\{i[H,c_{d\uparrow}]\delta
j^\mp_{d\downarrow}, \delta
j^\pm_{d\downarrow}c_{d\uparrow}^\dagger\}\rangle
=-i\epsilon_d\langle\{c_{d\uparrow}\delta j^\mp_{d\downarrow},
\delta j^\pm_{d\downarrow}c_{d\uparrow}^\dagger\}\rangle
-iU\langle\{c_{d\uparrow}n_{d\downarrow}\delta
j^\mp_{d\downarrow}, \delta
j^\pm_{d\downarrow}c_{d\uparrow}^\dagger\}\rangle-i\sum_{\bf
k}V^*_{{\bf k}d}\langle\{c_{k\uparrow}\delta j^\mp_{d\downarrow},
\delta j^\pm_{d\downarrow}c_{d\uparrow}^\dagger\}\rangle$. The
first and second terms of this expression should vanish because
they are not written in the form of a squared norm. Since the
third term describes the Kondo process, we treat it rigorously and
have the following expression: $-i\sum_{\bf k}V^*_{{\bf
k}d}\langle\{ c_{k\uparrow}\delta j^\mp_{d\downarrow}, \delta
j^\pm_{d\downarrow}c_{d\uparrow}^\dagger\}\rangle \\ =-i\sum_{\bf
k}V^*_{{\bf k}d}\langle\{ \delta j^\mp_{d\downarrow}\delta
j^\pm_{d\downarrow} c_{k\uparrow}c_{d\uparrow}^\dagger -\delta
j^\pm_{d\downarrow}\delta j^\mp_{d\downarrow}
c_{k\uparrow}c_{d\uparrow}^\dagger\}\rangle \\ =-i\sum_{\bf
k}V^*_{{\bf k}d}\langle[\delta j^\mp_{d\downarrow}, \delta
j^\pm_{d\downarrow}]c_{k\uparrow}c_{d\uparrow}^\dagger \rangle \\
=-i\sum_{\bf k}V^*_{{\bf k}d}\langle[j^\mp_{d\downarrow},
j^\pm_{d\downarrow}]c_{k\uparrow}c_{d\uparrow}^\dagger \rangle.$
The second term of $\langle\{i{\bf L}(c_{d\uparrow}\delta
j^\mp_{d\downarrow}),\delta
j^\pm_{d\downarrow}c_{d\uparrow}^\dagger\} \rangle$, i.e.,
$\langle\{c_{d\uparrow}i[H,j^\mp_{d\downarrow}], \delta
j^\pm_{d\downarrow}c_{d\uparrow}^\dagger\}\rangle$,
 is rather complicate because it contains the commutator
$i[H,j^\mp_{d\downarrow}]$. This commutator is expanded as
$i[H,j^\mp_{d\downarrow}]=\pm\epsilon_d
j^\pm_{d\downarrow}\mp(i)\sum_k
\epsilon_k(V_{kd}c_{k\downarrow}^\dagger c_{d\downarrow}\pm
V_{kd}^*c_{d\downarrow}^\dagger c_{k\downarrow})\pm
Un_{d\uparrow}j^\pm_{d\downarrow}\pm(i)
\sum_{k,\ell}^\prime(V_{kd}V_{\ell d}c_{k\downarrow}^\dagger
c_{\ell\downarrow}\pm V_{kd}^*V_{\ell
d}^*c_{\ell\downarrow}^\dagger c_{k\downarrow})$, where $(i)$ is
applied only to the lower signs and the prime in sum denotes
$k\neq\ell$. The first two terms on the right side are cancelled
if we assume $\epsilon_k\approx\epsilon_d$, and the $U$-term is
neglected in the Kondo regime. The last term, second order
hybridization, describes the round trip of a down-spin electron
between the Kondo impurity and the lead, which is equivalent to
$(j^\pm_{d\downarrow})^2$. Since we consider only single trip of a
down-spin electron in this study, we neglect the term involving
$[H,j^\mp_{d\downarrow}]$. Thus, the matrix element $({\bf
M}_{d})_{21/12}=\frac{-\langle\{iL(c_{d\uparrow}\delta
j^\mp_{d\downarrow}), \delta
j^\pm_{d\downarrow}c_{d\uparrow}^\dagger\}\rangle}{\sqrt{\langle(
\delta j^\mp_{d\downarrow})^2\rangle}\sqrt{\langle( \delta
j^\pm_{d\downarrow})^2\rangle}}=\pm\gamma$, where
$\gamma=\pm\frac{i\sum_{\bf k}V^*_{{\bf k}d}\langle
c_{k\uparrow}c_{d\uparrow}^\dagger[j^\mp_{d\downarrow},
j^\pm_{d\downarrow}] \rangle}{\sqrt{\langle( \delta
j^\mp_{d\downarrow})^2\rangle}\sqrt{\langle( \delta
j^\pm_{d\downarrow})^2\rangle}}$ for the single-reservoir Anderson
model. $\gamma_{LL(RR)}$ for the two-reservoir Anderson model is
given by
\[\gamma_{LL(RR)}=\frac{i\sum_{\bf k}V^*_{{\bf k}d}\langle c_{k\uparrow}^{L}
c^\dagger_{d\uparrow}+c_{k\uparrow}^{R}c^\dagger_{d\uparrow})
[j^{-L(R)}_{d\downarrow},j^{+L(R)}_{d\downarrow}]\rangle}{\langle
(\delta j^{-L(R)}_{d\downarrow})^2\rangle^{1/2}\langle (\delta
j^{+L(R)}_{d\downarrow})^2\rangle^{1/2}}.
\]

\section*{Appendix B: Expressions of $\beta_{mn}$}
We obtain the expressions of $\beta_{mn}$ in terms of the
definitions and the expressions of $\xi_i^{\mp}$ given in Eq.
(\ref{22}). The coefficients $\beta_{mn}$ in front of the
self-energy function, i.e.,
$i\Sigma_{mn}(\omega)=\beta_{mn}(i\Sigma^L_0(\omega)+i\Sigma^R_0(\omega))$,
are symmetric in exchanging their indices. By using the operator
identity $[n_{i\downarrow},j^\mp_{i\downarrow}]=\mp
ij^\pm_{i\downarrow}$, we write a different expression for ${\rm
Re}[U^{\pm L,R}_{J^\pm}]$, i.e., ${\rm Re}[U^{\pm
L,R}_{J^\pm}]\equiv(U/2\tau)$, where
$$1/\tau=[\pm\langle j^{\pm L,R}_{i\downarrow}\rangle\mp 2\langle j^{\pm
L,R}_{i\downarrow}n_{i\uparrow}\rangle]/\sqrt{\langle (\delta
j^{\mp L,R}_{i\downarrow})^2\rangle},$$ which will be used in the
expressions of ${\rm Re}[\beta_{mn}]$. The final forms of ${\rm
Re}[\beta_{mn}]$ are given by using $\langle(\delta
n_{i\downarrow})^2\rangle\approx 1/4$ and $\tau=2$ that is the
value at the atomic limit. Since $\langle
j_{i\downarrow}^{+L(R)}\rangle<0$ and $\langle
j_{i\downarrow}^{-L}\rangle=-\langle j_{i\downarrow}^{-R}\rangle$,
an inequality, ${\rm Re}[\beta_{12(14)}]<{\rm Re}[\beta_{15}]<{\rm
Re}[\beta_{11(55)}]<{\rm Re}[\beta_{25(45)}]<{\rm
Re}[\beta_{22(44)}]$, exists. We obtain the expressions of
$\beta_{mn}$ as follows:
\vspace{0.2cm} \\
\noindent {\bf Real parts of $\beta_{mn}$:}
\\ ${\rm Re}[\beta_{11}]={\rm Re}[\xi_L^{-\ast}\xi_L^{-}]=\\ \left[(\langle
j_{i\downarrow}^{+L}\rangle-2\langle
n_{i\uparrow}j_{i\downarrow}^{+L}\rangle)^2+(1-2\langle
n_{i\downarrow}\rangle)^2\langle
j_{i\downarrow}^{-L}\rangle^2\right] \frac{[4\langle(\delta
n_{i\downarrow})^2\rangle]^{-1}}{\langle (\delta
j_{i\downarrow}^{-L})^2\rangle} \\
=\left[\frac{1}{\tau^2}+\frac{(1-2\langle
n_{i\downarrow}\rangle)^2\langle
j_{i\downarrow}^{-L}\rangle^2}{\langle (\delta
j_{i\downarrow}^{-L})^2\rangle}\right]\frac{1}{4\langle(\delta
n_{i\downarrow})^2\rangle}\\
\approx\left[\frac{1}{4}+\frac{(1-2\langle
n_{i\downarrow}\rangle)^2\langle
j_{i\downarrow}^{-L}\rangle^2}{\langle (\delta
j_{i\downarrow}^{-L})^2\rangle}\right].$ \\Only the index $L$
changes to $R$ for ${\rm Re}[\beta_{55}]$.
\vspace{0.1cm} \\
${\rm Re}[\beta_{22}]={\rm Re}[\xi_L^{+\ast}\xi_L^{+}]=\\
\left[(\langle j_{i\downarrow}^{-L}\rangle-2\langle
n_{i\uparrow}j_{i\downarrow}^{-L}\rangle)^2+(1-2\langle
n_{i\downarrow}\rangle)^2\langle
j_{i\downarrow}^{+L}\rangle^2\right]\frac{[4\langle(\delta
n_{i\downarrow})^2\rangle]^{-1}}{\langle (\delta
j_{i\downarrow}^{+L})^2\rangle}\\=\left[\frac{1}{\tau^2}+\frac{(1-2\langle
n_{i\downarrow}\rangle)^2\langle
j_{i\downarrow}^{+L}\rangle^2}{\langle (\delta
j_{i\downarrow}^{+L})^2\rangle}\right]\frac{1}{4\langle(\delta
n_{i\downarrow})^2\rangle}
\\ \approx\left[\frac{1}{4}+\frac{(1-2\langle
n_{i\downarrow}\rangle)^2\langle
j_{i\downarrow}^{+L}\rangle^2}{\langle (\delta
j_{i\downarrow}^{+L})^2\rangle}\right]$.\\ Here, again, the index
$L$ changes to $R$ for ${\rm Re}[\beta_{44}]$.
\vspace{0.1cm} \\
${\rm Re}[\beta_{12}]={\rm Re}[\xi_L^{-\ast}\xi_L^{+}]={\rm
Re}[\beta_{21}]=\\\left[(\langle
j_{i\downarrow}^{+L}\rangle-2\langle
n_{i\uparrow}j_{i\downarrow}^{+L}\rangle)(-\langle
j_{i\downarrow}^{-L}\rangle+2\langle
n_{i\uparrow}j_{i\downarrow}^{-L}\rangle)\right. \\ \left.
+(1-2\langle n_{i\downarrow}\rangle)^2\langle
j_{i\downarrow}^{-L}\rangle\langle
j_{i\downarrow}^{+L}\rangle\right]\frac{[4\langle(\delta
n_{i\downarrow})^2\rangle]^{-1}}{\sqrt{\langle (\delta
j_{i\downarrow}^{-L})^2\rangle}\sqrt{\langle (\delta
j_{i\downarrow}^{+L})^2\rangle}} \\
=\left[\frac{1}{\tau^2}+\frac{(1-2\langle
n_{i\downarrow}\rangle)^2\langle
j_{i\downarrow}^{-L}\rangle\langle
j_{i\downarrow}^{+L}\rangle}{\sqrt{\langle (\delta
j_{i\downarrow}^{-L})^2\rangle}\sqrt{\langle (\delta
j_{i\downarrow}^{+L})^2\rangle}}\right]\frac{1}{4\langle(\delta
n_{i\downarrow})^2\rangle}\\
\approx\left[\frac{1}{4}+\frac{(1-2\langle
n_{i\downarrow}\rangle)^2\langle
j_{i\downarrow}^{-L}\rangle\langle
j_{i\downarrow}^{+L}\rangle}{\sqrt{\langle (\delta
j_{i\downarrow}^{-L})^2\rangle}\sqrt{\langle (\delta
j_{i\downarrow}^{+L})^2\rangle}}\right],$
\vspace{0.1cm} \\
${\rm Re}[\beta_{14}]={\rm Re}[\xi_L^{-\ast}\xi_R^{+}]={\rm
Re}[\beta_{41}] \vspace{0.1cm}=\\
\left[(\langle j_{i\downarrow}^{+L}\rangle-2\langle
n_{i\uparrow}j_{i\downarrow}^{+L}\rangle)(-\langle
j_{i\downarrow}^{-R}\rangle+2\langle
n_{i\uparrow}j_{i\downarrow}^{-R}\rangle)\right.
\\ \left.+(1-2\langle n_{i\downarrow}\rangle)^2\langle
j_{i\downarrow}^{-L}\rangle\langle
j_{i\downarrow}^{+R}\rangle\right]\frac{[4\langle(\delta
n_{i\downarrow})^2\rangle]^{-1}}{\sqrt{\langle (\delta
j_{i\downarrow}^{-L})^2\rangle}\sqrt{\langle (\delta
j_{i\downarrow}^{+R})^2\rangle}} \\
=\left[\frac{1}{\tau^2}+\frac{(1-2\langle
n_{i\downarrow}\rangle)^2\langle
j_{i\downarrow}^{-L}\rangle\langle
j_{i\downarrow}^{+R}\rangle}{\sqrt{\langle (\delta
j_{i\downarrow}^{-L})^2\rangle}\sqrt{\langle (\delta
j_{i\downarrow}^{+R})^2\rangle}}\right]\frac{1}{4\langle(\delta
n_{i\downarrow})^2\rangle}
\\ \approx\left[\frac{1}{4}+\frac{(1-2\langle
n_{i\downarrow}\rangle)^2\langle
j_{i\downarrow}^{-L}\rangle\langle
j_{i\downarrow}^{+R}\rangle}{\sqrt{\langle (\delta
j_{i\downarrow}^{-L})^2\rangle}\sqrt{\langle (\delta
j_{i\downarrow}^{+R})^2\rangle}}\right] ,$
\vspace{0.1cm} \\
${\rm Re}[\beta_{15}]={\rm Re}[\xi_L^{-\ast}\xi_R^{-}]={\rm
Re}[\beta_{51}] \vspace{0.1cm}=\\ \left[(\langle
j_{i\downarrow}^{+L}\rangle-2\langle
n_{i\uparrow}j_{i\downarrow}^{+L}\rangle)(\langle
j_{i\downarrow}^{+R}\rangle-2\langle
n_{i\uparrow}j_{i\downarrow}^{+R}\rangle)\right.\\
\left.+(1-2\langle n_{i\downarrow}\rangle)^2\langle
j_{i\downarrow}^{-L}\rangle\langle
j_{i\downarrow}^{-R}\rangle\right]\frac{[4\langle(\delta
n_{i\downarrow})^2\rangle]^{-1}}{\sqrt{\langle (\delta
j_{i\downarrow}^{-L})^2\rangle}\sqrt{\langle (\delta
j_{i\downarrow}^{-R})^2\rangle}} \\
=\left[\frac{1}{\tau^2}+\frac{(1-2\langle
n_{i\downarrow}\rangle)^2\langle
j_{i\downarrow}^{-L}\rangle\langle
j_{i\downarrow}^{-R}\rangle}{\sqrt{\langle (\delta
j_{i\downarrow}^{-L})^2\rangle}\sqrt{\langle (\delta
j_{i\downarrow}^{-R})^2\rangle}}\right]\frac{1}{4\langle(\delta
n_{i\downarrow})^2\rangle}\\
\approx\left[\frac{1}{4}+\frac{(1-2\langle
n_{i\downarrow}\rangle)^2\langle
j_{i\downarrow}^{-L}\rangle\langle
j_{i\downarrow}^{-R}\rangle}{\sqrt{\langle (\delta
j_{i\downarrow}^{-L})^2\rangle}\sqrt{\langle (\delta
j_{i\downarrow}^{-R})^2\rangle}}\right],$
\vspace{0.1cm} \\
${\rm Re}[\beta_{24}]={\rm Re}[\xi_L^{+\ast}\xi_R^{+}]={\rm
Re}[\beta_{42}] \vspace{0.1cm}=\\ \left[(2\langle
n_{i\uparrow}j_{i\downarrow}^{-L}\rangle-\langle
j_{i\downarrow}^{-L}\rangle)(2\langle
n_{i\uparrow}j_{i\downarrow}^{-R}\rangle-\langle
j_{i\downarrow}^{-R}\rangle)\right.\\ \left.+(1-2\langle
n_{i\downarrow}\rangle)^2\langle
j_{i\downarrow}^{+L}\rangle\langle
j_{i\downarrow}^{+R}\rangle\right]\frac{[4\langle(\delta
n_{i\downarrow})^2\rangle]^{-1}}{\sqrt{\langle (\delta
j_{i\downarrow}^{+L})^2\rangle}\sqrt{\langle (\delta
j_{i\downarrow}^{+R})^2\rangle}} \\
=\left[\frac{1}{\tau^2}+\frac{(1-2\langle
n_{i\downarrow}\rangle)^2\langle
j_{i\downarrow}^{+L}\rangle\langle
j_{i\downarrow}^{+R}\rangle}{\sqrt{\langle (\delta
j_{i\downarrow}^{+L})^2\rangle}\sqrt{\langle (\delta
j_{i\downarrow}^{+R})^2\rangle}}\right]\frac{1}{4\langle(\delta
n_{i\downarrow})^2\rangle}\\
\approx\left[\frac{1}{4}+\frac{(1-2\langle
n_{i\downarrow}\rangle)^2\langle
j_{i\downarrow}^{+L}\rangle\langle
j_{i\downarrow}^{+R}\rangle}{\sqrt{\langle (\delta
j_{i\downarrow}^{+L})^2\rangle}\sqrt{\langle (\delta
j_{i\downarrow}^{+R})^2\rangle}}\right], $\\
\vspace{0.1cm} \\
${\rm Re}[\beta_{25}]={\rm Re}[\xi_L^{+\ast}\xi_R^{-}]={\rm
Re}[\beta_{52}] \vspace{0.1cm}=\\ \left[(-\langle
j_{i\downarrow}^{-L}\rangle+2\langle
n_{i\uparrow}j_{i\downarrow}^{-L}\rangle)(\langle
j_{i\downarrow}^{+R}\rangle-2\langle
n_{i\uparrow}j_{i\downarrow}^{+R}\rangle)\right.\\
\left.+(1-2\langle n_{i\downarrow}\rangle)^2\langle
j_{i\downarrow}^{+L}\rangle\langle
j_{i\downarrow}^{-R}\rangle\right]\frac{[4\langle(\delta
n_{i\downarrow})^2\rangle]^{-1}}{\sqrt{\langle (\delta
j_{i\downarrow}^{+L})^2\rangle}\sqrt{\langle (\delta
j_{i\downarrow}^{-R})^2\rangle}} \\
=\left[\frac{1}{\tau^2}+\frac{(1-2\langle
n_{i\downarrow}\rangle)^2\langle
j_{i\downarrow}^{+L}\rangle\langle
j_{i\downarrow}^{-R}\rangle}{\sqrt{\langle (\delta
j_{i\downarrow}^{+L})^2\rangle}\sqrt{\langle (\delta
j_{i\downarrow}^{-R})^2\rangle}}\right]\frac{1}{4\langle(\delta
n_{i\downarrow})^2\rangle}\\
\approx\left[\frac{1}{4}+\frac{(1-2\langle
n_{i\downarrow}\rangle)^2\langle
j_{i\downarrow}^{+L}\rangle\langle
j_{i\downarrow}^{-R}\rangle}{\sqrt{\langle (\delta
j_{i\downarrow}^{+L})^2\rangle}\sqrt{\langle (\delta
j_{i\downarrow}^{-R})^2\rangle}}\right],$ \\
\vspace{0.1cm} \\
${\rm Re}[\beta_{45}]={\rm Re}[\xi_R^{+\ast}\xi_R^{-}]={\rm
Re}[\beta_{54}] \vspace{0.1cm}=\\ \left[(-\langle
j_{i\downarrow}^{-R}\rangle+2\langle
n_{i\uparrow}j_{i\downarrow}^{-R}\rangle)(\langle
j_{i\downarrow}^{+R}\rangle-2\langle
n_{i\uparrow}j_{i\downarrow}^{+R}\rangle)\right. \\
\left.+(1-2\langle n_{i\downarrow}\rangle)^2\langle
j_{i\downarrow}^{+R}\rangle\langle
j_{i\downarrow}^{-R}\rangle\right]\frac{[4\langle(\delta
n_{i\downarrow})^2\rangle]^{-1}}{\sqrt{\langle (\delta
j_{i\downarrow}^{+R})^2\rangle}\sqrt{\langle (\delta
j_{i\downarrow}^{-R})^2\rangle}} \\
=\left[\frac{1}{\tau^2}+\frac{(1-2\langle
n_{i\downarrow}\rangle)^2\langle
j_{i\downarrow}^{+R}\rangle\langle
j_{i\downarrow}^{-R}\rangle}{\sqrt{\langle (\delta
j_{i\downarrow}^{+R})^2\rangle}\sqrt{\langle (\delta
j_{i\downarrow}^{-R})^2\rangle}}\right]\frac{1}{4\langle(\delta
n_{i\downarrow})^2\rangle}\\
\approx\left[\frac{1}{4}+\frac{(1-2\langle
n_{i\downarrow}\rangle)^2\langle
j_{i\downarrow}^{+R}\rangle\langle
j_{i\downarrow}^{-R}\rangle}{\sqrt{\langle (\delta
j_{i\downarrow}^{+R})^2\rangle}\sqrt{\langle (\delta
j_{i\downarrow}^{-R})^2\rangle}}\right]$.
\vspace{0.2cm} \\

\noindent {\bf Imaginary parts of $\beta_{mn}$:}\\
\noindent ${\rm Im}[\beta_{12}]={\rm
Im}[\xi_{L}^{-*}\xi_{L}^{+}]\\
=\frac{(1-2\langle n_{i\downarrow}\rangle)}{4\langle(\delta
n_{i\downarrow})^2\rangle}\left[\frac{\langle
j_{i\downarrow}^{+L}\rangle-2\langle
n_{i\uparrow}j_{i\downarrow}^{+L}\rangle}{ \sqrt{\langle (\delta
j_{i\downarrow}^{-L})^2\rangle}} \frac{\langle
j_{i\downarrow}^{+L}\rangle}{\sqrt{\langle (\delta
j_{i\downarrow}^{+L})^2\rangle}}\right. \\ \left. -\frac{-\langle
j_{i\downarrow}^{-L}\rangle+2\langle
n_{i\uparrow}j_{i\downarrow}^{-L}\rangle}{\sqrt{\langle (\delta
j_{i\downarrow}^{+L})^2\rangle}}\frac{\langle
j_{i\downarrow}^{-L}\rangle}{\sqrt{\langle (\delta
j_{i\downarrow}^{-L})^2\rangle}}\right]\\
=\frac{(1-2\langle n_{i\downarrow}\rangle)}{4\langle(\delta
n_{i\downarrow})^2\rangle}\frac{1}{\tau}\left[\frac{\langle
j_{i\downarrow}^{+L}\rangle}{\sqrt{\langle (\delta
j_{i\downarrow}^{+L})^2\rangle}}-\frac{\langle
j_{i\downarrow}^{-L}\rangle}{\sqrt{\langle (\delta
j_{i\downarrow}^{-L})^2\rangle}}\right] $

\noindent ${\rm Im}[\beta_{14}]={\rm
Im}[\xi_{L}^{-*}\xi_{R}^{+}]\\
=\frac{(1-2\langle n_{i\downarrow}\rangle)}{4\langle(\delta
n_{i\downarrow})^2\rangle}\left[\frac{\langle
j_{i\downarrow}^{+L}\rangle-2\langle
n_{i\uparrow}j_{i\downarrow}^{+L}\rangle}{ \sqrt{\langle (\delta
j_{i\downarrow}^{-L})^2\rangle}} \frac{\langle
j_{i\downarrow}^{+R}\rangle}{\sqrt{\langle (\delta
j_{i\downarrow}^{+R})^2\rangle}}\right. \\ \left. -\frac{-\langle
j_{i\downarrow}^{-R}\rangle+2\langle
n_{i\uparrow}j_{i\downarrow}^{-R}\rangle}{\sqrt{\langle (\delta
j_{i\downarrow}^{+R})^2\rangle}}\frac{\langle
j_{i\downarrow}^{-L}\rangle}{\sqrt{\langle (\delta
j_{i\downarrow}^{-L})^2\rangle}}\right]\\
=\frac{(1-2\langle n_{i\downarrow}\rangle)}{4\langle(\delta
n_{i\downarrow})^2\rangle}\frac{1}{\tau}\left[\frac{\langle
j_{i\downarrow}^{+R}\rangle}{\sqrt{\langle (\delta
j_{i\downarrow}^{+R})^2\rangle}}-\frac{\langle
j_{i\downarrow}^{-L}\rangle}{\sqrt{\langle (\delta
j_{i\downarrow}^{-L})^2\rangle}}\right] $

\noindent ${\rm Im}[\beta_{15}]={\rm Im}[\xi_{L}^{-*}\xi_{R}^{-}] \\
=\frac{(1-2\langle n_{i\downarrow}\rangle)}{4\langle(\delta
n_{i\downarrow})^2\rangle}\left[\frac{\langle
j_{i\downarrow}^{+L}\rangle-2\langle
n_{i\uparrow}j_{i\downarrow}^{+L}\rangle}{ \sqrt{\langle (\delta
j_{i\downarrow}^{-L})^2\rangle}} \frac{\langle
j_{i\downarrow}^{-R}\rangle}{\sqrt{\langle (\delta
j_{i\downarrow}^{-R})^2\rangle}}\right.\\ \left. -\frac{\langle
j_{i\downarrow}^{+R}\rangle-2\langle
n_{i\uparrow}j_{i\downarrow}^{+R}\rangle}{\sqrt{\langle (\delta
j_{i\downarrow}^{-R})^2\rangle}}\frac{\langle
j_{i\downarrow}^{-L}\rangle}{\sqrt{\langle (\delta
j_{i\downarrow}^{-L})^2\rangle}}\right]\\
=\frac{(1-2\langle n_{i\downarrow}\rangle)}{4\langle(\delta
n_{i\downarrow})^2\rangle}\frac{1}{\tau}\left[\frac{\langle
j_{i\downarrow}^{-R}\rangle}{\sqrt{\langle (\delta
j_{i\downarrow}^{-R})^2\rangle}}-\frac{\langle
j_{i\downarrow}^{-L}\rangle}{\sqrt{\langle (\delta
j_{i\downarrow}^{-L})^2\rangle}}\right] $,

\noindent ${\rm Im}[\beta_{25}]=\frac{(1-2\langle
n_{i\downarrow}\rangle)}{4\langle(\delta
n_{i\downarrow})^2\rangle}\left[\frac{-\langle
j_{i\downarrow}^{-L}\rangle+2\langle
n_{i\uparrow}j_{i\downarrow}^{-L}\rangle}{ \sqrt{\langle (\delta
j_{i\downarrow}^{+L})^2\rangle}} \frac{\langle
j_{i\downarrow}^{-R}\rangle}{\sqrt{\langle (\delta
j_{i\downarrow}^{-R})^2\rangle}}\right. \\ \left. -\frac{\langle
j_{i\downarrow}^{+R}\rangle-2\langle
n_{i\uparrow}j_{i\downarrow}^{+R}\rangle}{\sqrt{\langle (\delta
j_{i\downarrow}^{-R})^2\rangle}}\frac{\langle
j_{i\downarrow}^{+L}\rangle}{\sqrt{\langle (\delta
j_{i\downarrow}^{+L})^2\rangle}}\right]\\
=\frac{(1-2\langle n_{i\downarrow}\rangle)}{4\langle(\delta
n_{i\downarrow})^2\rangle}\frac{1}{\tau}\left[\frac{\langle
j_{i\downarrow}^{-R}\rangle}{\sqrt{\langle (\delta
j_{i\downarrow}^{-R})^2\rangle}}-\frac{\langle
j_{i\downarrow}^{+L}\rangle}{\sqrt{\langle (\delta
j_{i\downarrow}^{+L})^2\rangle}}\right] $,

\noindent ${\rm Im}[\beta_{45}]=\frac{(1-2\langle
n_{i\downarrow}\rangle)}{4\langle(\delta
n_{i\downarrow})^2\rangle}\left[\frac{-\langle
j_{i\downarrow}^{-R}\rangle+2\langle
n_{i\uparrow}j_{i\downarrow}^{-R}\rangle}{ \sqrt{\langle (\delta
j_{i\downarrow}^{+R})^2\rangle}} \frac{\langle
j_{i\downarrow}^{-R}\rangle}{\sqrt{\langle (\delta
j_{i\downarrow}^{-R})^2\rangle}}\right. \\ \left. -\frac{\langle
j_{i\downarrow}^{+R}\rangle-2\langle
n_{i\uparrow}j_{i\downarrow}^{+R}\rangle}{\sqrt{\langle (\delta
j_{i\downarrow}^{-R})^2\rangle}}\frac{\langle
j_{i\downarrow}^{+R}\rangle}{\sqrt{\langle (\delta
j_{i\downarrow}^{+R})^2\rangle}}\right]\\
=\frac{(1-2\langle n_{i\downarrow}\rangle)}{4\langle(\delta
n_{i\downarrow})^2\rangle}\frac{1}{\tau}\left[\frac{\langle
j_{i\downarrow}^{-R}\rangle}{\sqrt{\langle (\delta
j_{i\downarrow}^{-R})^2\rangle}}-\frac{\langle
j_{i\downarrow}^{+R}\rangle}{\sqrt{\langle (\delta
j_{i\downarrow}^{+R})^2\rangle}}\right] $, and

\noindent ${\rm Im}[\beta_{24}]=\frac{(1-2\langle
n_{i\downarrow}\rangle)}{4\langle(\delta
n_{i\downarrow})^2\rangle}\left[\frac{-\langle
j_{i\downarrow}^{-L}\rangle+2\langle
n_{i\uparrow}j_{i\downarrow}^{-L}\rangle}{ \sqrt{\langle (\delta
j_{i\downarrow}^{+L})^2\rangle}} \frac{\langle
j_{i\downarrow}^{+R}\rangle}{\sqrt{\langle (\delta
j_{i\downarrow}^{+R})^2\rangle}}\right.\\ \left. -\frac{-\langle
j_{i\downarrow}^{-R}\rangle+2\langle
n_{i\uparrow}j_{i\downarrow}^{-R}\rangle}{\sqrt{\langle (\delta
j_{i\downarrow}^{+R})^2\rangle}}\frac{\langle
j_{i\downarrow}^{+L}\rangle}{\sqrt{\langle (\delta
j_{i\downarrow}^{+L})^2\rangle}}\right]\\
=\frac{(1-2\langle n_{i\downarrow}\rangle)}{4\langle(\delta
n_{i\downarrow})^2\rangle}\frac{1}{\tau}\left[\frac{\langle
j_{i\downarrow}^{+R}\rangle}{\sqrt{\langle (\delta
j_{i\downarrow}^{+R})^2\rangle}}-\frac{\langle
j_{i\downarrow}^{+L}\rangle}{\sqrt{\langle (\delta
j_{i\downarrow}^{+L})^2\rangle}}\right]$.


\end{document}